\documentclass[runningheads]{llncs}
\usepackage{graphicx}
\usepackage{bm}
\usepackage{amsfonts}
\usepackage{mathrsfs} 
\usepackage{amsmath}
\usepackage[super]{nth}
\usepackage{url}
\usepackage{perpage} 
\MakePerPage{footnote}
\usepackage{graphicx}
\usepackage{booktabs,multirow}
\usepackage{arydshln}
\usepackage{colortbl} 
\usepackage{xcolor}
\usepackage{color}
\definecolor{blue1}{RGB}{221,235,247}
\definecolor{yellow1}{RGB}{255,242,204}
\definecolor{green1}{RGB}{226,239,218}
\definecolor{pink}{RGB}{255,239,236}
\usepackage{hyperref}
\usepackage[misc]{ifsym}
\begin{document}
\title{Beyond the Snapshot: Brain Tokenized Graph Transformer for Longitudinal Brain Functional Connectome Embedding}
\titlerunning{Brain TokenGT}
%
\author{Zijian Dong\inst{1,2} \and
Yilei Wu\inst{1} \and
Yu Xiao\inst{1} \and Joanna Su Xian Chong\inst{1} \and Yueming Jin\inst{4,2} \and Juan Helen Zhou\inst{1,2,3}\textsuperscript{(\Letter)}}
%
\authorrunning{Z.Dong et al.}
%
\institute{Centre for Sleep and Cognition \& Centre for Translational Magnetic Resonance Research, Yong Loo Lin School of Medicine, National University of Singapore, Singapore \and Department of Electrical and Computer Engineering, National University of Singapore, Singapore \and Integrative Sciences and Engineering Programme (ISEP), NUS Graduate School, National University of Singapore, Singapore \and Department of Biomedical Engineering, National University of Singapore, Singapore \\ \email{helen.zhou@nus.edu.sg}}
\maketitle              
%
\begin{abstract}
Under the framework of network-based neurodegeneration, brain functional connectome (FC)-based Graph Neural Networks (GNN) have emerged as a valuable tool for the diagnosis and prognosis of neurodegenerative diseases such as Alzheimer's disease (AD). However, these models are tailored for brain FC at a single time point instead of characterizing FC trajectory. Discerning how FC evolves with disease progression, particularly at the predementia stages such as cognitively normal individuals with amyloid deposition or individuals with mild cognitive impairment (MCI), is crucial for delineating disease spreading patterns and developing effective strategies to slow down or even halt disease advancement. In this work, we proposed the \emph{first} interpretable framework for brain FC trajectory embedding with application to neurodegenerative disease diagnosis and prognosis, namely \emph{Brain Tokenized Graph Transformer} (Brain TokenGT). It consists of two modules: 1)  \emph{Graph Invariant and Variant Embedding} (GIVE) for generation of node and spatio-temporal edge embeddings, which were tokenized for downstream processing; 2) \emph{Brain Informed Graph Transformer Readout} (BIGTR) which augments previous tokens with trainable type identifiers and non-trainable node identifiers and feeds them into a standard transformer encoder to readout. We conducted extensive experiments on two public longitudinal fMRI datasets of the AD continuum for three tasks, including differentiating MCI from controls, predicting dementia conversion in MCI, and classification of amyloid positive or negative cognitively normal individuals. Based on brain FC trajectory, the proposed Brain TokenGT approach outperformed all the other benchmark models and at the same time provided excellent interpretability. 

\end{abstract}
\section{Introduction}
The brain functional connectome (FC) is a graph with brain regions of interest (ROIs) represented as nodes and pairwise correlations of fMRI time series between the ROIs represented as edges. FC has been shown to be a promising biomarker for the early diagnosis and tracking of neurodegenerative disease progression (\emph{e.g.}, Alzheimer's Disease (AD)) because of its ability to capture disease-related alternations in brain functional organization \cite{zhou2012predicting,zhou2017applications}. Recently, the graph neural networks (GNN) has become the model of choice for processing graph structured data with state-of-the-art performance in different tasks \cite{corso2020principal,kipfsemi,velivckovicgraph}. With regards to FC, GNN has also shown promising results in disease diagnosis \cite{cui2022braingb,cui2022interpretable,kanbrain,li2021braingnn,zhang2021deep}. However, such studies have only focused on FC at a single time point. For neurodegenerative diseases like AD, it is crucial to investigate longitudinal FC changes \cite{filippi2020changes}, including graph topology \emph{and} attributes, in order to slow down or even halt disease advancement. 

Node features are commonly utilized in FC to extract important information. It is also essential to recognize the significance of edge features in FC, which are highly informative in characterizing the interdependencies between ROIs. Furthermore, node embeddings obtained from GNN manipulation contain essential information that should be effectively leveraged. Current GNNs feasible to graphs with multiple time points \cite{pareja2020evolvegcn,yu2018spatio,zheng2019onionnet} are suboptimal to FC trajectory, as they fail to incorporate brain edge feature embeddings and/or they rely on conventional operation (\emph{e.g.}, global pooling for readout) which introduces inductive bias and is incapable of extracting sufficient information from the node embeddings \cite{ying2018hierarchical}. Moreover, these models lack built-in interpretability, which is crucial for clinical applications. And they are unsuitable for small-scale datasets which are common in fMRI research. The longitudinal data with multiple time points of the AD continuum is even more scarce due to the difficulty in data acquisition.

\begin{figure}
\includegraphics[width=\textwidth]{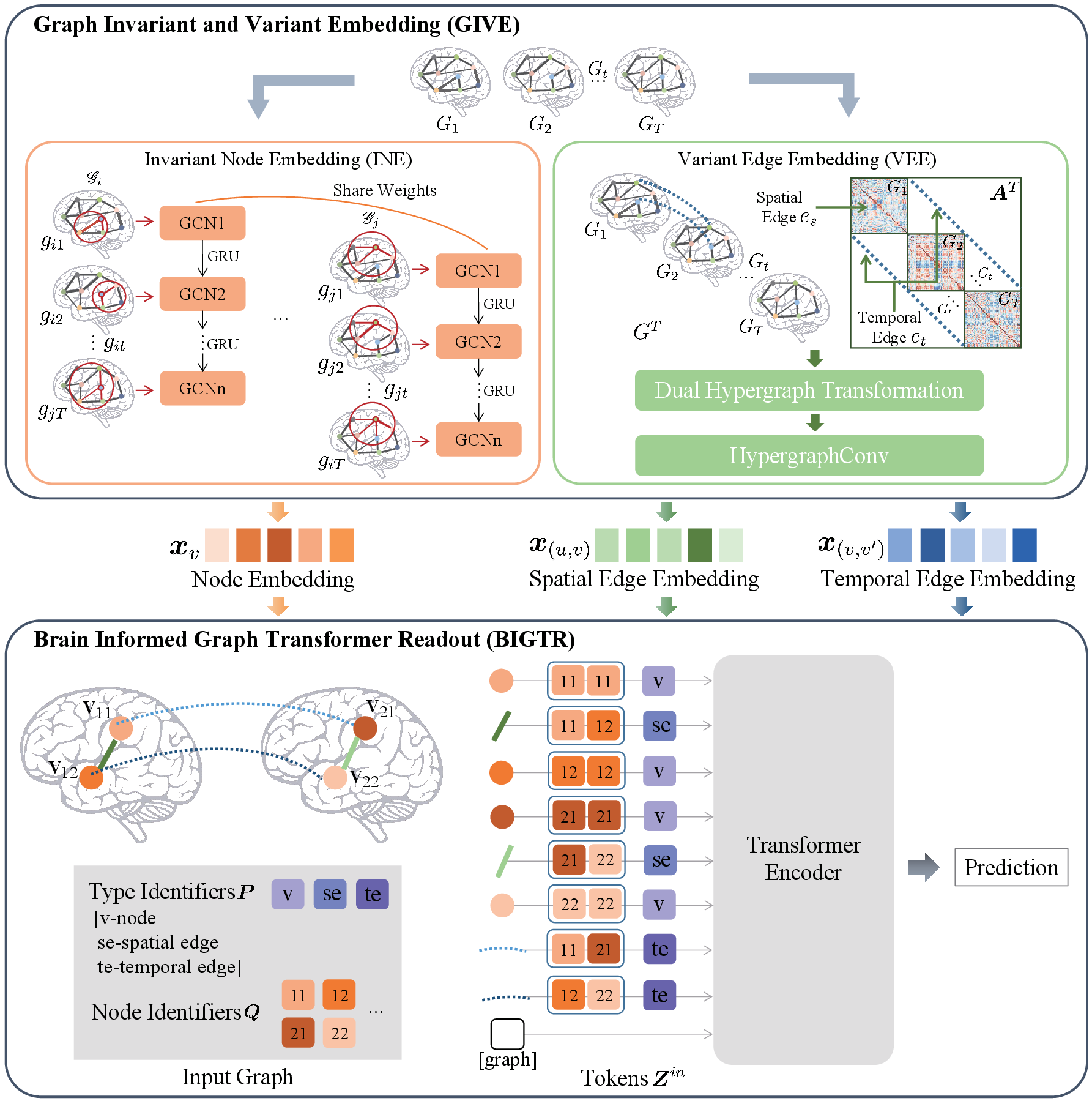}
\caption{An overview of Brain TokenGT. In GIVE, INE generates node embedding by performing evolving convolution on dynamic neighbourhood graph, and VEE combines different time points by defining spatio-temporal edge, and then transforms the whole trajectory into a dual hypergraph and produces spatial and temporal edge embedding. These embeddings, augmented by trainable type identifiers and non-trainable node identifiers, are used as input to a standard transformer encoder for readout within BIGTR.} \label{fig1}
\end{figure}

In this work, we proposed \emph{Brain Tokenized Graph Transformer} (Brain TokenGT), the \emph{first} framework to achieve FC trajectory embeddings with built-in interpretability, shown in Fig.~\ref{fig1}. Our contributions are as follows: 1) Drawing on the distinctive characteristics of FC trajectories, we developed \emph{Graph Invariant and Variant Embedding} (GIVE), which is capable of generating embeddings for both nodes and spatio-temporal edges; 2) Treating embeddings from GIVE as tokens, \emph{Brain Informed Graph Transformer Readout} (BIGTR) augments tokens with trainable type identifiers and non-trainable node identifiers and feeds them into a standard transformer encoder to readout instead of global pooling, further extracting information from tokens and alleviating over-fitting issue by token-level task; 3) We conducted extensive experiments on two public resting state fMRI datasets (ADNI, OASIS) with three different tasks (Healthy Control (HC) vs. Mild Cognition Impairment (MCI) classification, AD conversion prediction and Amyloid positive vs. negative classification). Our model showed superior results with FC trajectory as input, accompanied by node and edge level interpretations.

\section{Method}

\subsection{Problem Definition}
The input of one subject to the proposed framework is a sequence of brain networks $\mathcal{G}=[G_{1},G_{2},...,G_{t},...,G_{T}]$ with $T$ time points. Each network is a graph $G=(V,E,\bm{A})$, with the node set $V=\{v_{i}\}_{i=1}^{M}$, the edge set $E = V \times V$, and the weighted adjacency matrix $\bm{A} \in \mathbb{R}^{M \times M}$ describing the degrees of FC between ROIs. The output of the model is an individual-level categorical diagnosis $\hat{y}_{s}$ for each subject $s$. 

\subsection{Graph Invariant and Variant Embedding (GIVE)}

Regarding graph topology, one of the unique characteristics of FC across a trajectory is that it has invariant number and sequence of nodes (ROIs), with variant connections between different ROIs. Here, we designed GIVE, which consists of Invariant Node Embedding (INE) and Variant Edge Embedding (VEE).  

\subsubsection{Invariant Node Embedding (INE).} To obtain node embeddings that capture the spatial and temporal information of the FC trajectory, we utilized evolving graph convolution \cite{pareja2020evolvegcn} for the K-hop neighbourhood around each node which could be seen as a fully dynamic graph, providing a novel "zoom in" perspective to see FC. As suggested in \cite{pareja2020evolvegcn}, with informative node features, we chose to treat parameters in graph convolutional layers as hidden states of the dynamic system and used a gated recurrent unit (GRU) to update the hidden states.

Formally, for each node $v_{i}$ in $V$, we define a dynamic neighbourhood graph as $\mathscr{G}_{i}=[g_{i1},g_{i2},..,g_{it},...,g_{iT}]$ (Fig.~\ref{fig1}), in which $g_{it}$ is the K-hop neighbourhood of node $v_{i}$ at time point $t$, with adjacency matrix $\bm{A}_{it}$. At time $t$, for dynamic neighbourhood graph $\mathscr{G}_{i}$, $l$-th layer of evolving graph convolution first updates parameter matrix ${\bm{W}^{l}_{i(t-1)}}$ from the last time point to ${\bm{W}^{l}_{it}}$ with GRU, then the node embeddings ${\bm{H}^{l}_{it}}$ are updated to ${\bm{H}^{l+1}_{it}}$ for next layer using graph convolution network (GCN) \cite{kipfsemi}:
\begin{equation}
{\bm{W}^{l}_{it}}=\text{GRU}({\bm{H}^{l}_{it}},{\bm{W}^{l}_{i(t-1)}}); \quad {\bm{H}^{l+1}_{it}}=\text{GCN}(\bm{A}_{it},{\bm{H}^{l}_{it}},{\bm{W}^{l}_{it}})
\end{equation}

\subsubsection{Variant Edge Embedding (VEE).}For tasks such as graph classification, an appropriate representation of edges also plays a key role in the successful graph representation learning. To achieve edge embeddings, we first integrated graphs from multiple time points by defining \emph{Spatial Edge} and \emph{Temporal Edge}, and then obtained spatial and temporal edge embeddings by transforming an FC trajectory to the dual hypergraph. 

For each FC trajectory, we should not only investigate the edges between different ROIs in one static FC (\emph{i.e.}, spatial domain) but also capture the longitudinal change across different time points (\emph{i.e.}, time domain). Instead of focusing only on intrinsic connections (\emph{i.e.}, \emph{spatial edges} ($e_{s}$)) between different ROIs in each FC, for each of the two consecutive graphs $G_{t}$ and $G_{t+1}$, we added $M$ \emph{temporal edges} ($e_{t}$) to connect corresponding nodes in $G_{t}$ and $G_{t+1}$, with weights initialized as 1. The attached features to spatial and temporal edges were both initialized by the concatenation of node features from both ends and their initial weights. 

Accordingly, one trajectory would be treated as a single graph for downstream edge embedding. We denote the giant graph with $T$ time points contained as $G^{T}$, with weighted adjacency matrix $\bm{A}^{T} \in \mathbb{R}^{TM \times TM}$ (Fig.~\ref{fig1}). $G^{T}$ was first transformed into the dual hypergraph $G^{T*}$ by Dual Hypergraph Transformation (DHT) \cite{jo2021edge}, where the role of nodes and edges in $G^{T}$ was exchanged while their information was preserved. DHT is accomplished by transposing the incidence matrix of the graph to the new incidence matrix of the dual graph, which is formally defined as: $G^{T}=(\bm{X},\bm{M},\bm{E}) \mapsto G^{T*}=(\bm{E},\bm{M}^{\bm{T}},\bm{X})$, where $\bm{X} \in \mathbb{R}^{M \times D}$ is the original node features matrix with a $D$ dimensional feature vector for each node, $\bm{M} \in \mathbb{R}^{|E| \times M}$ is the original incidence matrix, and $\bm{E} \in \mathbb{R}^{|E| \times (2D+1)}$ is the initialized edge features matrix.

We then performed hypergraph convolution \cite{bai2021hypergraph} to achieve node embeddings in $G^{T*}$, which were the corresponding edge embeddings in $G^{T}$. The hypergraph convolution at $l^{th}$ layer is defined by:
\begin{equation}
\bm{E}^{(l+1)}=\bm{D}^{-1}\bm{M}^{\bm{T}}\bm{W^{*}}\bm{B}^{-1}\bm{M}\bm{E}^{(l)}\bm{\Theta}
\end{equation}
where $\bm{W^{*}}$ is the diagonal hyperedge weight matrix, $\bm{D}$ and $\bm{B}$ are the degree matrices of the nodes and hyperedges respectively, and $\bm{\Theta}$ is the parameters matrix.

Interpretability is important in decision-critical areas (\emph{e.g.}, disorder analysis). Thanks to the design of spatio-temporal edges, we could achieve built-in binary level interpretability (\emph{i.e.}, both nodes and edges contributing most to the given task, from $e_{t}$ and $e_{s}$, respectively) by leveraging HyperDrop \cite{jo2021edge}. The HyperDrop procedure is defined as follows: 
\begin{equation}
\text{idx}=\text{TopE}(\text{score}(\bm{E})); \quad \bm{E}^{\text{pool}}=\bm{E}_{\text{idx}}; \quad (\bm{M}^{\text{pool}})^{\bm{T}}=(\bm{M}^{\bm{T}}_{\text{idx}})
\end{equation}
where 'score' function is hypergraph convolution layers used to compute scores for each hypergraph node ($e_{s}$ or $e_{t}$ in the original graph). 'TopE' selects the nodes with the highest E scores (note: ranking was performed for nodes from $e_{s}$ and $e_{t}$ separately, and HyperDrop was only applied to nodes from $e_{s}$ with hyperparameter E), and idx is the node-wise indexing vector. Finally, the salient nodes (from $e_{t}$) and edges (from $e_{s}$) were determined by ranking the scores averaged across the trajectory.

\subsection{Brain Informed Graph Transformer Readout (BIGTR)}
Proper readout for the embeddings from GNN manipulation is essential to produce meaningful prediction outcome for assisting diganosis and prognosis. The vanilla ways are feeding the Node Embeddings, and Spatial and Temporal Edge Embeddings generated from the GIVE module into pooling and fully connected layers. However, this would result in a substantial loss of spatial and temporal information \cite{ying2018hierarchical}, especially under the complex settings of three types of spatial/temporal embeddings. Recently, it has been shown, both in theory and practice, that a standard transformer with appropriate token embeddings yields a powerful graph learner \cite{kimpure}. Here, treating embeddings output from GIVE as tokens, we leveraged graph transformer as a trainable readout function, named as Brain Informed Graph Transformer Readout (BIGTR) (Fig.~\ref{fig1}).

We first define the Type Identifier (TI) and Node Identifier (NI) under the setting of FC trajectory. \emph{Trainable TI} encodes whether a token is a node, spatial edge or temporal edge. They are defined as a parameter matrix $[\bm{P}_{v};\bm{P}_{e_{s}};\bm{P}_{e_{t}}] \in \mathbb{R}^{3 \times d_{p}}$, where $\bm{P}_{v}$, $\bm{P}_{{e}_{s}}$ and $\bm{P}_{{e}_{t}}$ are node, spatial edge and temporal edge identifier respectively. Specifically, we maintained a dictionary, in which the keys are types of the tokens, the values are learnable embeddings that encodes the corresponding token types. It facilitates the model's learning of type-specific attributes in tokens, compelling attention heads to focus on disease-related token disparities, thereby alleviating overfitting caused by non-disease-related attributes. Besides, it inflates 1 $G^{T}$ for an individual-level task to thousands of tokens, which could also alleviate overfitting in the perspective of small-scale datasets. \emph{Non-trainable NI} are $MT$ node-wise orthonormal vectors $\bm{Q} \in \mathbb{R}^{MT \times d_{q}}$ for an FC trajectory with $T$ time points and $M$ nodes at each time. Then, the augmented token features become:
\begin{equation}
\begin{aligned}
\bm{z}_{v}&=[\bm{x}_v,\bm{P}_{v},\bm{Q}_{v},\bm{Q}_{v}] \\ \bm{z}_{(u,v)}&=[\bm{x}_{(u,v)},\bm{P}_{e_{s}},\bm{Q}_{u},\bm{Q}_{v}] \\ \bm{z}_{(v,v')}&=[\bm{x}_{(v,v')},\bm{P}_{e_{t}},\bm{Q}_{v},\bm{Q}_{v']}
\end{aligned}
\end{equation}
for $v$, $e_{s}$ and $e_{t}$ respectively, where node $u$ is a neighbour to node $v$ in the spatial domain and node $v'$ is a neighbour to node $v$ in the temporal domain, and $\bm{x}$ is the original token from GIVE. Thus, the augmented token features matrix is $\bm{Z} \in \mathbb{R}^{(MT+|E|T+M(T-1)) \times (h+d_{p}+2d_{q})}$, where $h$ is the hidden dimension of embeddings from GIVE. $\bm{Z}$ would be further projected by a trainable matrix $\omega \in \mathbb{R}^{(h+d_{p}+2d_{q}) \times h'}$. As we targeted individual-level (\emph{i.e.}, $G^{T}$) diagnosis/prognosis, a graph token $\bm{X}_{[\text{graph}]} \in \mathbb{R}^{h'}$ was appended as well. Thus, the input to transformer is formally defined as : 
\begin{equation}
\bm{Z}^{in}=[\bm{X}_{[\text{graph}]};\bm{Z}\omega] \in \mathbb{R}^{(1+MT+|E|T+M(T-1)) \times h'}
\end{equation}

\section{Experiments}

\subsubsection{Datasets and Experimental settings.} We used brain FC metrics derived from ADNI \cite{jack2008alzheimer} and OASIS-3 \cite{lamontagne2019oasis} resting state fMRI datasets, following preprocessing pipelines \cite{kong2019spatial,li2019global}. Our framework was evaluated on three classification tasks related to diagnosis or prognosis: 1)  HC vs. MCI classification (ADNI: 65 HC \& 60 MCI), 2) AD conversion prediction (OASIS-3: 31 MCI non-converters \& 29 MCI converters), and 3) differentiating cognitively normal individuals with amyloid positivity vs. those with amyloid negativity (OASIS-3: 41 HC a$\beta$+ve \& 50 HC a$\beta$-ve). All subjects have 2-3 time points of fMRI data and those with two time points were zero-padded to three time points. FC was built based on the AAL brain atlas with 90 ROIs \cite{tzourio2002automated}. The model was trained using Binary Cross-Entropy Loss in an end-to-end fashion. Implementation details could be found in supplementary materials. The code is available at \url{https://github.com/ZijianD/Brain-TokenGT.git}

\begin{table}[h]
	\centering
	\caption{Experimental Results reported based on five-fold cross-validation repeated five times (\%, mean(standard deviation)). Our approach outperformed shallow learning (in blue), one time point feasible deep learning (in yellow), multi-time point feasible deep learning (in green) and our ablations (in pink) significantly. [* denotes significant improvement ($p<0.05$). HC: healthy control. MCI: mild cognitive impairment. AD: Alzheimer's disease. GP: global pooling. I: identifiers. TI: type identifiers. NI: node identifiers.]}
	\scriptsize 
	\renewcommand\arraystretch{1.3}
	\tabcolsep=0.1mm
	\resizebox{\textwidth}{!}{
	\begin{tabular}{@{}ccccccccc@{}}
		\specialrule{1pt}{0pt}{0pt}
		\multirow{1}*{Model} & \multicolumn{2}{c}{\multirow{1}*{HC vs. MCI}}    && \multicolumn{2}{c}{\multirow{1}*{AD Conversion}} & & \multicolumn{2}{c}{Amyloid Positive vs. Negative} \\
			\cdashline{2-3}[0.8pt/2pt]\cdashline{5-6}[0.8pt/2pt]\cdashline{8-9}[0.8pt/2pt]
		\multicolumn{1}{c}{} & {AUC} & {Accuracy} && {AUC} & {Accuracy} & &{AUC} & {Accuracy} \\
		\cline{1-1}
		\cellcolor{blue1}MK-SVM & {55.00{\tiny (15.31)}} & {47.20{\tiny (06.40)}} && {56.69{\tiny (14.53)}} & {58.19{\tiny (08.52)}} && {61.31{\tiny (11.16)}} & {56.02{\tiny (07.91)}} \\
		\cellcolor{blue1}RF    & {55.26{\tiny (13.83)}} & {57.60{\tiny (07.42)}} && {62.00{\tiny (03.46)}} & {56.44{\tiny (02.50)}} & &{65.25{\tiny (09.41)}} & {60.35{\tiny (08.48)}} \\
		\cellcolor{blue1}MLP   & {59.87{\tiny (12.89)}} & {51.20{\tiny (09.26)}} && {59.19{\tiny (13.76)}} & {58.13{\tiny (13.67)}} & &{60.33{\tiny (13.60)}} & {59.30{\tiny (13.52)}} \\
		\cline{2-3}\cline{5-6}\cline{8-9}
		\cellcolor{yellow1}GCN   & {62.86{\tiny (00.79)}} & {58.33{\tiny (01.03)}} && {62.22{\tiny (04.75)}} & {63.33{\tiny (03.78)}} && {66.67{\tiny (00.67)}} & {66.67{\tiny (00.87)}} \\
		\cellcolor{yellow1}GAT   & {61.11{\tiny (00.38)}} & {61.54{\tiny (01.47)}} && {63.83{\tiny (00.95)}} & {46.67{\tiny (07.11)}} && {68.00{\tiny (00.56)}} & {64.44{\tiny (00.69)}} \\
		\cellcolor{yellow1}PNA   & {65.25{\tiny (09.41)}} & {60.35{\tiny (08.48)}} && {67.11{\tiny (16.22)}} & {62.57{\tiny (08.38)}} && {70.08{\tiny (12.41)}} & {62.63{\tiny (05.42)}} \\
		\cellcolor{yellow1}BrainNetCNN & {48.06{\tiny (05.72)}} & {55.73{\tiny (06.82)}} && {60.79{\tiny (09.12)}} & {58.33{\tiny (11.65)}} && {65.80{\tiny (01.60)}} & {67.84{\tiny (02.71)}} \\
		\cellcolor{yellow1}BrainGNN & {60.94{\tiny (07.85)}} & {51.80{\tiny (07.49)}} && {69.38{\tiny (15.97)}} & {59.93{\tiny (12.30)}} && {62.40{\tiny (09.18)}} & {63.90{\tiny (10.75)}} \\
		\cellcolor{yellow1}IBGNN+ & {67.95{\tiny (07.95)}} & {66.79{\tiny (08.10)}} && {75.98{\tiny (06.36)}} & {70.00{\tiny (05.70)}} && {73.45{\tiny (05.61)}} & {65.28{\tiny (04.73)}} \\
		\cellcolor{yellow1}BrainNetTF & {65.03{\tiny (07.11)}} & {63.08{\tiny (07.28)}} && {73.33{\tiny (10.07)}} & {73.33{\tiny (08.16)}} && {74.32{\tiny (06.46)}} & {76.84{\tiny (06.67)}} \\
		\cline{2-3}\cline{5-6}\cline{8-9}
		\cellcolor{green1}OnionNet & {61.52{\tiny (06.16)}} & {60.20{\tiny (06.87)}} && {65.08{\tiny (03.69)}} & {63.89{\tiny (06.33)}} && {60.74{\tiny (08.47)}} & {64.38{\tiny (10.06)}} \\
		\cellcolor{green1}STGCN & {76.62{\tiny (06.77)}} & {74.77{\tiny (09.59)}} && {76.92{\tiny (06.89)}} & {75.00{\tiny (04.05)}} && {78.69{\tiny (03.59)}} & {75.03{\tiny (06.00)}} \\
		\cellcolor{green1}EvolveGCN & {81.52{\tiny (02.48)}} & {80.45{\tiny (02.52)}} && {77.63{\tiny (02.78)}} & {76.99{\tiny (06.53)}} && {79.33{\tiny (03.34)}} & {75.14{\tiny (09.59)}} \\
            \cline{2-3}\cline{5-6}\cline{8-9}
		\cellcolor{pink}GIVE w/o $\rm{e_t}$ $+$ GP & {82.83{\tiny (09.50)}} & {83.97{\tiny (04.76)}} && {78.14{\tiny (02.37)}} & {80.02{\tiny (06.62)}} && {80.35{\tiny (06.09)}} & {75.34{\tiny (09.76)}} \\
		\cellcolor{pink}GIVE $+$ GP & {83.11{\tiny (05.66)}} & {85.50{\tiny (01.47)}} && {79.21{\tiny (05.50)}} & {81.45{\tiny (04.58)}} && {83.57{\tiny (08.07)}} & {80.85{\tiny (03.67)}} \\
		\cellcolor{pink}BIGTR itself & {76.83{\tiny (00.67)}} & {71.33{\tiny (00.87)}} && {71.17{\tiny (00.38)}} & {63.33{\tiny (01.47)}}& & {77.61{\tiny (01.17)}} & {74.00{\tiny (00.22)}} \\
		\cellcolor{pink}Ours w/o I & {85.17{\tiny (06.37)}} & {84.11{\tiny (03.20)}} && {83.80{\tiny (04.03)}} & {83.20{\tiny (02.40)}} && {84.24{\tiny (07.01)}} & {83.20{\tiny (02.40)}} \\
            \cellcolor{pink}Ours w/o TI & {86.71{\tiny (08.01)}} & {89.02{\tiny (04.37)}} && {85.17{\tiny (06.37)}} & {84.11{\tiny (03.20)}} && {88.40{\tiny (06.95)}} & {84.67{\tiny (05.64)}} \\
            \cellcolor{pink}Ours w/o NI & {88.15{\tiny (08.07)}} & {\textbf{91.04{\tiny (06.15)}}} && {84.17{\tiny (03.17)}} & {85.80{\tiny (07.19)}} && {88.70{\tiny (03.98)}} & {84.36{\tiny (04.32)}} \\
		\cellcolor{pink}\textbf{Ours [GIVE + BIGTR]}  & {\textbf{90.48*{\tiny (04.99)}}} & {84.62{\tiny (08.43)}} && {\textbf{87.14*{\tiny (07.16)}}} & {\textbf{89.23*{\tiny (07.84)}}} && {\textbf{94.60*{\tiny (04.96)}}} & {\textbf{87.11*{\tiny (07.88)}}} \\
		\specialrule{1pt}{0pt}{0pt}
	\end{tabular}}%
	\label{tab:01}%
\end{table}%

\subsubsection{Results.} AUC and accuracy are presented in Table \ref{tab:01}. (Recall and Precision could be found in supplementary materials). Brain TokenGT and its ablations were compared with three types of baseline models, including 1) shallow machine learning: MK-SVM, RF and MLP; 2) one time point feasible deep learning: three representative deep graph models GCN \cite{kipfsemi}, GAT \cite{velivckovicgraph} and PNA \cite{corso2020principal}, and four state-of-the-art deep models specifically designed for FC: BrainnetCNN \cite{kawahara2017brainnetcnn}, BrainGNN \cite{li2021braingnn}, IBGNN+ \cite{cui2022interpretable} and BrainnetTF \cite{kanbrain}; 3) multiple time points feasible deep learning: Onionnet \cite{zheng2019onionnet}, STGCN \cite{yu2018spatio} and EvolveGCN \cite{pareja2020evolvegcn}. To ensure a fair comparison between models, the one-dimensional vectors flattened from FC in all time points were concatenated and used as input for the shallow learning model. For the one time point feasible deep learning models, a prediction value was generated at each time point and subsequently averaged to obtain an individual-level prediction.

The experimental results (Table \ref{tab:01}) demonstrate that the Brain TokenGT significantly outperformed all three types of baseline by a large margin. The ablation study further revealed that GIVE w/o $e_{t}$ w/ GP outperformed EvolveGCN by adding VEE without $e_{t}$, which empirically validates the importance of edge feature embeddings in FC. The performance could be further improved by incorporating $e_{t}$, suggesting the efficiency of our GIVE design with spatio-temporal edges. Interestingly, BIGTR itself (\emph{i.e.}, the original features were directly input to BIGTR without GIVE) showed competitive performance with STGCN. Replacing GP with transformer (Ours w/o I) led to improved performance even without identifiers, indicating that the embeddings from GIVE may already capture some spatial and temporal information from the FC trajectory. The addition of identifiers further improved performance, possibly because the token-level self-supervised learning could alleviate the over-fitting issue and node identifiers could maintain the localized information effectively.

\subsubsection{Interpretation.}

\begin{figure}
\includegraphics[width=\textwidth]{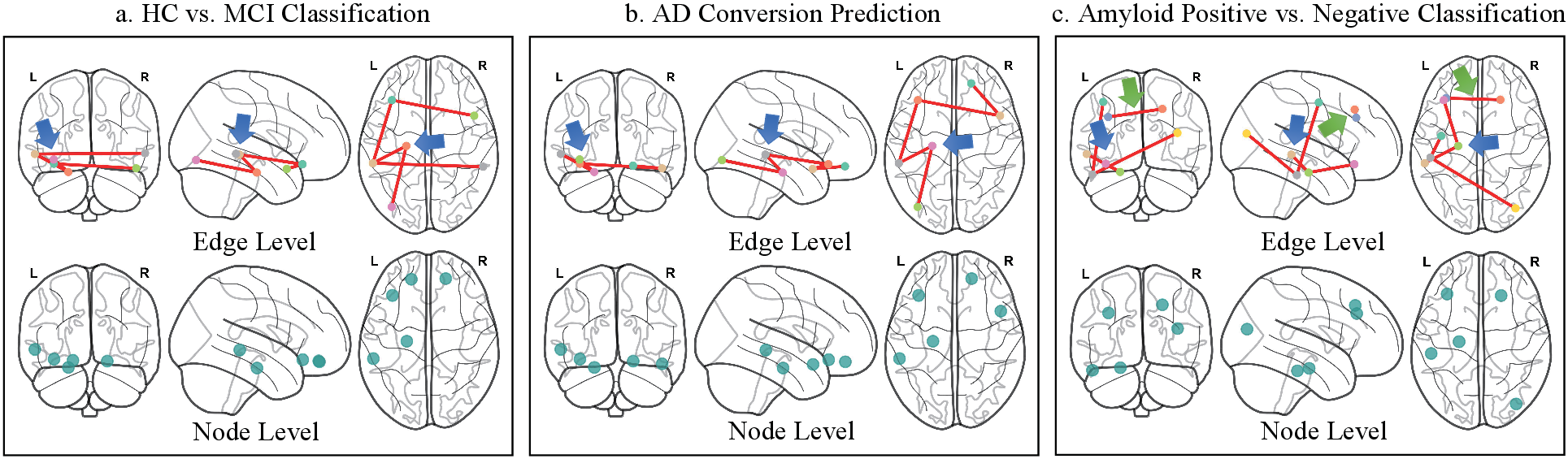}
\caption{HyperDrop Results. Blue arrows point to left temporal and parahippocampal regions, green arrows point to superior frontal regions. We refer readers of interest to supplementary materials for the full list of brain regions and connections involved.} \label{fig2}
\end{figure}

Fig.~\ref{fig2} shows the top 5 salient edges and nodes retained by HyperDrop for each of the three tasks. Consistent with previous literature on brain network breakdown in the early stage of AD \cite{sheline2013resting}, parahippocampal, orbitofrontal and temporal regions and their connections contributed highly to all three tasks, underscoring their critical roles in AD-specific network dysfunction relevant to disease progression. On the other hand, superior frontal region additionally contributed to the amyloid positive vs. negative classification, which is in line with previous studies in amyloid deposition \cite{thal2002phases}. 

\section{Conclusion}

This study proposes the first interpretable framework for the embedding of FC trajectories, which can be applied to the diagnosis and prognosis of neurodegenerative diseases with small scale datasets, namely \emph{Brain Tokenized Graph Transformer} (Brain TokenGT). Based on longitudinal brain FC, experimental results showed superior performance of our framework with excellent built-in interpretability supporting the AD-specific brain network neurodegeneration. A potential avenue for future research stemming from this study involves enhancing the "temporal resolution" of the model. This may entail, for example, incorporating an estimation of uncertainty in both diagnosis and prognosis, accounting for disease progression, and offering time-specific node and edge level interpretation.

\section*{Acknowledgement}

This work was supported by National Medical Research Council, Singapore (NMRC/OFLCG19May-0035 to J-H Zhou) and Yong Loo Lin School of Medicine Research Core Funding (to J-H Zhou), National University of Singapore, Singapore. Yueming Jin was supported by MoE Tier 1 Start up grant (WBS: A-8001267-00-00).


%
%
%
\bibliographystyle{splncs04}
\bibliography{references}

\begin{thebibliography}{10}
\providecommand{\url}[1]{\texttt{#1}}
\providecommand{\urlprefix}{URL }
\providecommand{\doi}[1]{https://doi.org/#1}

\bibitem{bai2021hypergraph}
Bai, S., Zhang, F., Torr, P.H.: Hypergraph convolution and hypergraph
  attention. Pattern Recognition  \textbf{110},  107637 (2021)

\bibitem{corso2020principal}
Corso, G., Cavalleri, L., Beaini, D., Li{\`o}, P., Veli{\v{c}}kovi{\'c}, P.:
  Principal neighbourhood aggregation for graph nets. Advances in Neural
  Information Processing Systems  \textbf{33},  13260--13271 (2020)

\bibitem{cui2022braingb}
Cui, H., Dai, W., Zhu, Y., Kan, X., Gu, A.A.C., Lukemire, J., Zhan, L., He, L.,
  Guo, Y., Yang, C.: Braingb: a benchmark for brain network analysis with graph
  neural networks. IEEE Transactions on Medical Imaging  (2022)

\bibitem{cui2022interpretable}
Cui, H., Dai, W., Zhu, Y., Li, X., He, L., Yang, C.: Interpretable graph neural
  networks for connectome-based brain disorder analysis. In: Medical Image
  Computing and Computer Assisted Intervention--MICCAI 2022: 25th International
  Conference, Singapore, September 18--22, 2022, Proceedings, Part VIII. pp.
  375--385. Springer (2022)

\bibitem{filippi2020changes}
Filippi, M., Basaia, S., Canu, E., Imperiale, F., Magnani, G., Falautano, M.,
  Comi, G., Falini, A., Agosta, F.: Changes in functional and structural brain
  connectome along the alzheimer’s disease continuum. Molecular psychiatry
  \textbf{25}(1),  230--239 (2020)

\bibitem{jack2008alzheimer}
Jack~Jr, C.R., Bernstein, M.A., Fox, N.C., Thompson, P., Alexander, G., Harvey,
  D., Borowski, B., Britson, P.J., L.~Whitwell, J., Ward, C., et~al.: The
  alzheimer's disease neuroimaging initiative (adni): Mri methods. Journal of
  Magnetic Resonance Imaging: An Official Journal of the International Society
  for Magnetic Resonance in Medicine  \textbf{27}(4),  685--691 (2008)

\bibitem{jo2021edge}
Jo, J., Baek, J., Lee, S., Kim, D., Kang, M., Hwang, S.J.: Edge representation
  learning with hypergraphs. Advances in Neural Information Processing Systems
  \textbf{34},  7534--7546 (2021)

\bibitem{kanbrain}
Kan, X., Dai, W., Cui, H., Zhang, Z., Guo, Y., Yang, C.: Brain network
  transformer. In: Advances in Neural Information Processing Systems

\bibitem{kawahara2017brainnetcnn}
Kawahara, J., Brown, C.J., Miller, S.P., Booth, B.G., Chau, V., Grunau, R.E.,
  Zwicker, J.G., Hamarneh, G.: Brainnetcnn: Convolutional neural networks for
  brain networks; towards predicting neurodevelopment. NeuroImage
  \textbf{146},  1038--1049 (2017)

\bibitem{kimpure}
Kim, J., Nguyen, D.T., Min, S., Cho, S., Lee, M., Lee, H., Hong, S.: Pure
  transformers are powerful graph learners. In: Advances in Neural Information
  Processing Systems

\bibitem{kipfsemi}
Kipf, T.N., Welling, M.: Semi-supervised classification with graph
  convolutional networks. In: International Conference on Learning
  Representations

\bibitem{kong2019spatial}
Kong, R., Li, J., Orban, C., Sabuncu, M.R., Liu, H., Schaefer, A., Sun, N.,
  Zuo, X.N., Holmes, A.J., Eickhoff, S.B., et~al.: Spatial topography of
  individual-specific cortical networks predicts human cognition, personality,
  and emotion. Cerebral cortex  \textbf{29}(6),  2533--2551 (2019)

\bibitem{lamontagne2019oasis}
LaMontagne, P.J., Benzinger, T.L., Morris, J.C., Keefe, S., Hornbeck, R.,
  Xiong, C., Grant, E., Hassenstab, J., Moulder, K., Vlassenko, A.G., et~al.:
  Oasis-3: longitudinal neuroimaging, clinical, and cognitive dataset for
  normal aging and alzheimer disease. MedRxiv pp. 2019--12 (2019)

\bibitem{li2019global}
Li, J., Kong, R., Li{\'e}geois, R., Orban, C., Tan, Y., Sun, N., Holmes, A.J.,
  Sabuncu, M.R., Ge, T., Yeo, B.T.: Global signal regression strengthens
  association between resting-state functional connectivity and behavior.
  NeuroImage  \textbf{196},  126--141 (2019)

\bibitem{li2021braingnn}
Li, X., Zhou, Y., Dvornek, N., Zhang, M., Gao, S., Zhuang, J., Scheinost, D.,
  Staib, L.H., Ventola, P., Duncan, J.S.: Braingnn: Interpretable brain graph
  neural network for fmri analysis. Medical Image Analysis  \textbf{74},
  102233 (2021)

\bibitem{pareja2020evolvegcn}
Pareja, A., Domeniconi, G., Chen, J., Ma, T., Suzumura, T., Kanezashi, H.,
  Kaler, T., Schardl, T., Leiserson, C.: Evolvegcn: Evolving graph
  convolutional networks for dynamic graphs. In: Proceedings of the AAAI
  conference on artificial intelligence. vol.~34, pp. 5363--5370 (2020)

\bibitem{sheline2013resting}
Sheline, Y.I., Raichle, M.E.: Resting state functional connectivity in
  preclinical alzheimer’s disease. Biological psychiatry  \textbf{74}(5),
  340--347 (2013)

\bibitem{thal2002phases}
Thal, D.R., R{\"u}b, U., Orantes, M., Braak, H.: Phases of a$\beta$-deposition
  in the human brain and its relevance for the development of ad. Neurology
  \textbf{58}(12),  1791--1800 (2002)

\bibitem{tzourio2002automated}
Tzourio-Mazoyer, N., Landeau, B., Papathanassiou, D., Crivello, F., Etard, O.,
  Delcroix, N., Mazoyer, B., Joliot, M.: Automated anatomical labeling of
  activations in spm using a macroscopic anatomical parcellation of the mni mri
  single-subject brain. Neuroimage  \textbf{15}(1),  273--289 (2002)

\bibitem{velivckovicgraph}
Veli{\v{c}}kovi{\'c}, P., Cucurull, G., Casanova, A., Romero, A., Li{\`o}, P.,
  Bengio, Y.: Graph attention networks. In: International Conference on
  Learning Representations

\bibitem{ying2018hierarchical}
Ying, Z., You, J., Morris, C., Ren, X., Hamilton, W., Leskovec, J.:
  Hierarchical graph representation learning with differentiable pooling.
  Advances in neural information processing systems  \textbf{31} (2018)

\bibitem{yu2018spatio}
Yu, B., Yin, H., Zhu, Z.: Spatio-temporal graph convolutional networks: a deep
  learning framework for traffic forecasting. In: Proceedings of the 27th
  International Joint Conference on Artificial Intelligence. pp. 3634--3640
  (2018)

\bibitem{zhang2021deep}
Zhang, L., Wang, L., Gao, J., Risacher, S.L., Yan, J., Li, G., Liu, T., Zhu,
  D., Initiative, A.D.N., et~al.: Deep fusion of brain structure-function in
  mild cognitive impairment. Medical image analysis  \textbf{72},  102082
  (2021)

\bibitem{zheng2019onionnet}
Zheng, L., Fan, J., Mu, Y.: Onionnet: a multiple-layer
  intermolecular-contact-based convolutional neural network for protein--ligand
  binding affinity prediction. ACS omega  \textbf{4}(14),  15956--15965 (2019)

\bibitem{zhou2012predicting}
Zhou, J., Gennatas, E.D., Kramer, J.H., Miller, B.L., Seeley, W.W.: Predicting
  regional neurodegeneration from the healthy brain functional connectome.
  Neuron  \textbf{73}(6),  1216--1227 (2012)

\bibitem{zhou2017applications}
Zhou, J., Liu, S., Ng, K.K., Wang, J.: Applications of resting-state functional
  connectivity to neurodegenerative disease. Neuroimaging Clinics
  \textbf{27}(4),  663--683 (2017)

\end{thebibliography}

\end{document}